\documentclass[usegraphicx,usenatbib]{mnras}

\usepackage{newtxtext}

\DeclareRobustCommand{\VAN}[3]{#2}
\let\VANthebibliography\thebibliography
\def\thebibliography{\DeclareRobustCommand{\VAN}[3]{##3}\VANthebibliography}

\usepackage{amssymb}
\usepackage{dirtytalk}
\usepackage{lmodern}
\usepackage{caption}
\usepackage{threeparttable}
\usepackage[dvipsnames]{xcolor}
\usepackage{ulem}      
\usepackage[T1]{fontenc}
\usepackage{nicefrac}

\usepackage{graphicx}	
\usepackage{amsmath}	

\graphicspath{{./}{Figures/}}

\newcommand{\koral}{\texttt{KORAL}}
\newcommand{\Snew}{\texttt{S04}}
\newcommand{\uint}{u_\mathrm{int}}
\newcommand{\rg}{r_\mathrm{g}}

\newcommand{\tg}{t_\mathrm{g}}
\newcommand{\pg}{p_\mathrm{g}}
\newcommand{\pth}{p_\mathrm{th}}
\newcommand{\pmag}{p_\mathrm{m}}
\newcommand{\pr}{p_\mathrm{r}}
\newcommand{\mdot}{\dot{m}}
\newcommand{\dif}{\mathrm{d}}
\newcommand{\msol}{\mathrm{M}_\odot}
\newcommand{\orcid}[1]{\href{https://orcid.org/#1}{\includegraphics[width=8pt]{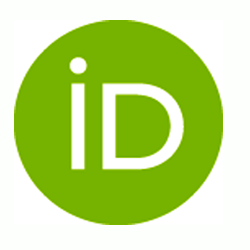}}}

\title[Radiative GRMHD simulations of puffy discs]{Radiative GRMHD simulations of puffy accretion discs:\\Numerical versus analytical models of sub-Eddington accretion}

\author[D. Lan{\v c}ov\'a]{Debora Lan{\v c}ov\'a\orcid{0000-0003-0826-9787}$^{1,2}$\thanks{E-mail: debora.lancova@gmail.com},
Maciek Wielgus\orcid{0000-0002-8635-4242}$^{3}$\thanks{E-mail: maciek@wielgus.info},
Marek Abramowicz\orcid{0000-0003-0067-5895}$^{1,2,4}$,
Agata R\'{o}\.{z}a\'{n}ska\orcid{0000-0002-5275-4096}$^{2}$,
\newauthor
W{\l}odek Klu{\'z}niak\orcid{0000-0001-9043-8062}$^{2}$,
Jiří Horák\orcid{0000-0002-7635-4839}$^{5}$,
David Abarca\orcid{0000-0002-9202-8734}$^{2}$,
Aleksander S{\k a}dowski$^{6}$, and Gabriel T\"{o}r\"{o}k\orcid{0000-0003-3958-9441}$^{1}$
\\
$^{1}$Research Center for Computational Physics and Data Processing, Institute of Physics, Silesian University in Opava,\\ Bezru\v{c}ovo n\'{a}m. 13, 746 01 Opava, Czech Republic\\
$^{2}$Nicolaus Copernicus Astronomical Centre of the Polish Academy of Sciences, Bartycka 18, 00-716 Warsaw, Poland\\
$^{3}$Instituto de Astrofísica de Andalucía-CSIC, Glorieta de la Astronomía s/n, E-18008 Granada, Spain\\
$^{4}$Department of Physics, G{\"o}teborg University, Origov\"{a}gen 6 B, 412 96 G{\"o}teborg, Sweden\\
$^{5}$Astronomical Institute of the Czech Academy of Sciences, Bo\v{c}n\'{\i} II 1401, CZ-14100 Prague, Czech Republic\\
$^{6}$Akuna Capital, 333 South Wabash Av., Chicago, IL 60604, USA}

\date{Accepted XXX. Received YYY; in original form ZZZ}

\pubyear{2026}

\begin{document}
\label{firstpage}
\pagerange{\pageref{firstpage}--\pageref{lastpage}}
\maketitle

\begin{abstract}
A widely accepted picture of an accretion flow in the luminous soft spectral state of X-ray binary systems is a geometrically thin disc structure much like the classic analytic solution of Shakura \& Sunyaev. Although the analytic models are troubled by instabilities and miss important aspects of physics, such as magnetic fields, they are successfully used as a framework for interpreting observational data. 
Here, we compare the results of general relativistic radiative magnetohydrodynamic (GRRMHD) simulations of optically thick, mildly sub-Eddington accretion on a stellar-mass black hole (the puffy disc) with established analytic and semi-analytic accretion models in the same regime. From the simulations, we find that the accretion flow is stabilised by the magnetic field, with a puffed-up, optically thick region resembling a warm corona surrounding a denser and cooler disc core. However, the stratified vertical structure of the disc significantly influences the observational picture of such a system. We analyse the inner disc structure, flow properties, effective viscosity, and inner edge position, and compare them to the predictions of standard models. We find that the simulated discs share some similarities with the models; however, they differ in several important aspects, most notably: the photosphere is geometrically thick, the inner edge is located closer to the central black hole than the analytic models assume, the surface density is significantly lower than analytically predicted, and the effective viscosity parameter is not constant but rises steeply in the innermost region.

\end{abstract}




\section{Introduction}

Since the beginning of X-ray observations, hundreds of X-ray binaries (XRBs) have been discovered, representing a unique laboratory for testing the behaviour of matter under extreme conditions, such as a~strong gravitational field, very high temperatures, and a~strong magnetic field. The material flowing from a companion forms an accretion disc around the compact object, which we can model using various analytic and numerical techniques and compare the models with the observed properties of XRBs. 

XRBs containing a black hole (BH) undergo irregular cycles of a faint quiescent state and luminous outbursts, during which radio emission from a jet is often observed. Thus, these systems are referred to as microquasars \citep{1992Natur.358..215M}. During these outbursts, the systems follow a clear transition path from hard to soft spectral states, going through an intermediate state in between \citep{2011soria}. State transitions are driven by changes in the mass accretion rate, which alter the disc structure and consequently the observed spectral and timing properties.

The spectral states during an outburst are well-defined: in the hard state, the hard X-ray luminosity is high and the spectrum is power-law dominated; in the soft state, the soft X-ray luminosity is high and the spectrum is thermal. The hard state is most often modelled by an optically thin inner disc transitioning into a standard thin disc at larger radii. The soft state is commonly envisioned as a standard Keplerian disc extending to the innermost stable circular orbit (ISCO) and radiating a multi-blackbody spectrum \citep{Done2007}.

Although the observable properties of XRBs during outbursts are determined by the state of the accretion disc, a comprehensive disc model that explains the observed behaviour self-consistently has not yet been proposed \citep{2004MNRAS.347..885G}. The thin disc model \citep{Shakura1973, Novikov1973}, most commonly used to fit the observed data, is unstable when the radiation pressure dominates, as shown analytically \citep{thermal76,viscous74,Piran1978},
and confirmed in numerical simulations \citep{2013ApJ...778...65J,Mishra2019,2023ApJ...945...57H}. However, in analytic and numerical modelling, the magnetic field was suggested and confirmed to be the most probable stabilising mechanism in the disc \citep[e.g.,][]{1995ApJ...445L..43M,2007MNRAS.375.1070B,Oda2009,2013MNRAS.434.2262Z,Sadowski2016,2020MNRAS.492.1855M,2022ApJ...939...31M,2023ApJ...954..150H,2024ApJ...966...47L,2025ApJ...995...26Z}. See also the review by \citet{2025arXiv250504402B} and references therein.

Another discrepancy between the observations and the thin disc model emerged just recently, with the polarimetric observations of XRBs using the Imaging X-ray Polarimetry Explorer \citep[IXPE, ][]{2022JATIS...8b6002W}. For example, observations of \texttt{4U 1630–47} in the soft state showed that current accretion disc models do not explain the high polarisation degree, despite providing a good spectral fit \citep{2024ApJ...964...77R}. 

Based on the work of \cite{Sadowski2016}, we performed simulations of a mildly sub-Eddington accretion disc to study its stability, properties, and observational signatures \citep{Lancova2019,Wielgus2022,2023AN....34430023L}. In contrast to the standard models of such a regime, we study a disc composed of material inflowing from a distant sub-Keplerian torus. The final disc, although sub-Eddington and supported radially by Keplerian rotation, remains thermally and viscously stable for a long time; however, it is no longer geometrically thin in the sense of the photospheric height and has a distinct vertical structure consisting of a dense core and a thick, opaque region surrounding it. Thus, we named this numerical solution the \textit{puffy accretion disc}. The puffy disc corresponds to the most luminous phase of a microquasar outburst, the thermal-component-dominated soft state.

In this work, we study the vertical structure and other properties of the puffy discs and compare them with standard analytic models of thin and slim discs. We build on previously published papers, introducing the stabilising mechanism \citep{Sadowski2016}, the unique structure \citep{Lancova2019}, and the consequences for observational signatures \citep{Wielgus2022,2023AN....34430023L}. We describe the inner vertical structure of the puffy disc and compare it with analytic models to show that although the puffy disc has some properties of a slim or thin disc, the overall structure is very different and reflects the complex behaviour of plasma under the influence of radiation and magnetic field. 
We compare simulations of four different values of mass accretion rates, with their properties summarised in \autoref{tab:sims}, but we focus mainly on simulation \texttt{S06}.

\renewcommand{\sectionautorefname}{Section}
\renewcommand{\subsectionautorefname}{Subsection}

The paper is organised as follows: in \autoref{sec:models}, we summarise the properties and current state of the analytic models of accretion. In \autoref{sec:koral}, we describe the \koral{} code and the simulation setup. \autoref{sec:puffy} discusses the general properties of the puffy disc, while \autoref{sec:alpha} focuses on the effective viscosity. Finally, \autoref{sec:conclusions} includes discussion and conclusions.

\section{Models of black hole accretion}
\label{sec:models}

\subsection{ {The thin disc}}

BH accretion is most often modelled by the standard thin accretion disc introduced by \citet{Shakura1973}, extended into general relativity (GR) by \citet{Novikov1973}. The model describes an optically thick, geometrically thin, and radiatively efficient Keplerian flow onto a central compact object. One of the main model assumptions is that the stress vanishes at the ISCO, and between the ISCO and the BH horizon (the plunging region), the flow is ionised, not thermalised, and optically thin. This property is often used to estimate the BH spin from X-ray observations \citep[see, e.g., ][ for a review]{2006ApJ...652..518M,2014SSRv..183..295M,2019NatAs...3...41R}.
We used the relativistic model of \citet{Novikov1973} for comparison of an analytic \say{standard} thin disc solution with our simulations.

The assumption of zero stress at the ISCO is valid for a razor-thin disc but not for a disc of realistic thickness, as was shown in analytic and numerical modelling \citep[e.g.,][]{1974ApJ...191..507T,1980AcA....30....1J,1999ApJ...515L..73K,2000ApJ...528..161A,2000astro.ph..4129P, 2001ApJ...548..348H,2002ApJ...573..754K,2005ApJ...621..372D,2008ApJ...687L..25S,Sadowski2009,2010ApJ...711..959N,2010MNRAS.408..752P,2023MNRAS.521.2439M,2024arXiv241006200L}, see also \autoref{sec:inneredge}. Spin estimation methods may then produce high values of the central BH spin because they fail to properly account for the emission from the plunging region \citep{Wielgus2022,2024MNRAS.533L..83M}, and because the inferred disc temperature is uncertain due to the poorly constrained colour correction factor  and simplified disc atmosphere modelling \citep{2023MNRAS.525.1288Y}.
However, the observations, especially those of microquasars in outbursts, align well with the thermal spectrum of a thin disc that remains stable for a very long time \citep{2009ApJ...701L..83S,Straub2011}.

\subsection{ {The slim disc}}

The slim disc model of \citet{Abramowicz1988} was introduced to resolve the problem of thermal instability of the thin disc in the radiation-pressure-dominated regime. It accounts for the advective cooling, stabilising the disc at high mass accretion rates. \citet{Sadowski2009} revisited the model computing the vertical structure after the surface (vertically integrated) density and radial velocity were determined from a one-dimensional model. The inner edge of the slim disc is located at the sonic point, which may differ from the ISCO. We adopted this revised slim disc model for comparison with the simulations in this work.

\begin{figure}
    \centering
    \includegraphics[width=1\linewidth]{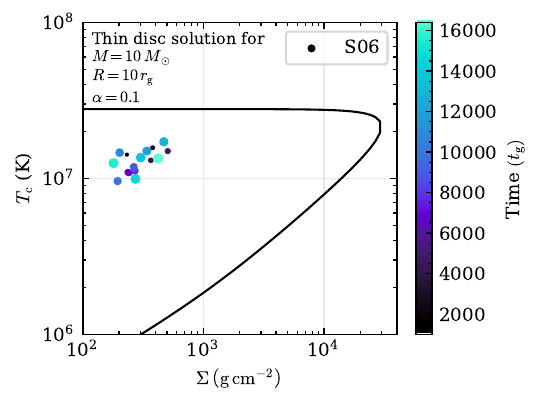}
    \caption{The surface density to central gas temperature ratio for the thin disc model (black curve) and simulation \texttt{S06} (see \autoref{tab:sims}) of the puffy disc at $R = 10\,\rg$. The size and colour of the dots correspond to different snapshots from the simulations, and time is shown on the colourbar. In the thin disc model, the upper constant part of the curve corresponds to the unstable radiation-pressure-dominated solution, and the bottom part to the stable gas-pressure-dominated branch.}
    \label{fig:scurve}
\end{figure}

\subsection{ {The puffy disc}}

Since a magnetic pressure-dominated discs, such as the puffy disc, are geometrically thicker than a thin disc, photons are more effectively trapped and advected with the flow, leading to stronger advective cooling than expected from the slim disc model at the same mass accretion rate. This is in accordance with the results of our simulations (see \autoref{sec:adv}). The vertical extension of a sub-Eddington accretion disc in a stable radiation-dominated regime was also shown in the results of several other simulations using various numerical codes and initial states \citep[see, e.g.,][]{2023MNRAS.518.3441C,2021ApJ...919L..20D,2024ApJ...966...47L,2022ApJ...939...31M,2023ApJ...959...59F,2023ApJ...945...57H,2025MNRAS.540.2820F,2026arXiv260305588Z}.

Numerical simulations of radiatively efficient accretion are computationally demanding due to the small geometrical thickness of the disc, which requires high resolution in the midplane. Another complication is the need to capture the cooling processes, e.g., using an artificial cooling function to remove all generated heat and keep the disc thin \citep{2008ApJ...687L..25S,2010MNRAS.408..752P} or to solve the coupling between gas and radiation. 
Still, in the radiation-pressure-dominated regime without the right magnetic field topology providing a radial or vertical net flux, the simulated disc collapses \citep{Sadowski2016, 2022ApJ...939...31M}.

The puffy disc simulations fall into the radiation-pressure-dominated regime, where the thin disc solution is unstable. Nevertheless, the sub-Eddington mass accretion rates discussed here are low for a typical slim disc application. Hence, the properties of the puffy disc place it between these two models, near the unstable branch of the S-curve \citep[][]{Abramowicz1988}. However, the simulated disc is stable, as illustrated in \autoref{fig:scurve}, which shows the surface density $\Sigma$ to central temperature $T_{\rm c}$ ratio for several snapshots from the puffy disc simulation and comparison with the analogous ratio computed for the analytic model. The ratio remains nearly constant throughout the simulation run. The $\Sigma$ of the simulated disc is lower than that of a stable thin disc while $T_{\rm c}$ is lower than that of the unstable horizontal branch due to magnetic pressure and advective cooling, as is further discussed in \autoref{sec:puffy}.

\section{Puffy disc simulations}
\label{sec:koral}

\subsection{The \koral{} code}

Here we summarise the \koral{} code details and the specific setup for the puffy disc problem. We perform the simulations using the \koral{} code\footnote{The latest version is available at \url{https://github.com/achael/koral_lite}}, a~highly parallelised general relativistic radiative magnetohydrodynamics (GRRMHD) code \citep{Sadowski2013, Sadowski2014, Sadowski2017}. \koral{} solves the set of relativistic ideal MHD equations of mass, energy, and momentum conservation\footnote{equations are given in $\mathrm{G=c}=1$ units, with Greek indices spanning four dimensions and Latin indices spanning the three spatial dimensions. We use the gravitational radius $\rg = \nicefrac{\mathrm{G}M}{\mathrm{c^2}}$ as the unit of length and $\tg = \nicefrac{\rg}{\mathrm{c}}$ as the unit of time.}.
\begin{align}
&\nabla_\mu \left(\rho u^\mu\right) = 0, \\
&\nabla_\mu \left( T^\mu\phantom{}_\nu + R^\mu\phantom{}_\nu \right) = 0, \label{eq:TplusR}
\end{align}
where $\rho$ is the gas rest mass density,
$u^\mu$ is the gas 4-velocity, and
$T^\mu\phantom{}_\nu$ and $R^\mu\phantom{}_\nu$ represent the MHD \citep[gas + magnetic field,][]{Gammie2003} stress-energy tensor, and the radiation stress-energy tensor, respectively. The MHD stress-energy tensor is given as
\begin{equation}
T^\mu\phantom{}_\nu\!= \!\left(\rho + u_{\mathrm{int}} + \pg + b^2\right)u^\mu u_\nu + 
\left(\pg+\dfrac{1}{2}b^2\right)\delta^\mu_\nu - b^\mu b_\nu \ , 
\end{equation}
where $\uint$ and $\pg$ are the gas internal energy and pressure measured in the comoving frame. They are related by the equation of state,
$\pg = (\gamma-1) \uint$ with assumed adiabatic index of polytropic gas $\gamma=\nicefrac{5}{3}$. The last expression in parentheses is related to the magnetic pressure as 
$\pmag = \nicefrac{b^2}{2}$, where the magnetic 4-vector $b^\mu$ is used, related to the magnetic field 3-vector $B^i$ \citep{Komissarov1999} by
\begin{align}
b^t &= B^i u^\mu g_{i\mu}, \\
b^i &= \dfrac{B^i + b^t u^i}{u^t},
\end{align}
for the spacetime metric $g_{\mu \nu}$.
The magnetic field is evolved using the relativistic 
induction equation
\begin{equation}
\partial _t \left(\sqrt{-g}B^i\right) = -\partial_j\left[\sqrt{-g}\left(b^j u^i - b^i u^j\right)\right],
\end{equation}
where $\sqrt{-g}$ is the metric determinant. The magnetic fluxes computed across the grid cells are calculated
using the flux-constrained transport algorithm of \citet{Toth2000} to preserve the divergence-free magnetic fields. 

\subsection{Radiation \texttt{M1} closure}

Radiation is evolved using the \texttt{M1} closure scheme \citep{Sadowski2013,Levermore1984}. It is worth noting here that this scheme is in some codes called the \textit{Levermore scheme}, while the \texttt{M1} refers to the scheme interpolating between the optically thin and thick regimes, such as in the \texttt{FRAC} code \citep{2020MNRAS.495.2285W}, see also \cite{2017MNRAS.469.1725M}. We will hold to the \texttt{M1} name in the context of the \koral{} code, as it was already used in many previous works. 

The \texttt{M1} closure, as implemented in \koral{}, assumes the existence of a frame with 4-velocity $u_{\mathrm{R}}^\mu$, in which the radiation is isotropic and characterised by the energy $\bar{E}$. The radiation tensor can then be constructed in any reference frame as

\begin{equation}
    R^{\mu\nu} = \dfrac{4}{3}\bar{E} u_{\mathrm{R}}^\mu u_{\mathrm{R}}^\nu + \dfrac{1}{3}
    \bar{E} g^{\mu \nu}.
\end{equation}

The coupling between gas and radiation is represented by a~radiation 4-force $G_\nu$ \citep{Mihalas1984}, so equation (\ref{eq:TplusR}) can be written as
\begin{align}
&\nabla_\mu T^\mu\phantom{}_\nu = G_\nu, \\
&\nabla _\mu R^\mu\phantom{}_\nu = -G_\nu.
\label{eq:consv}
\end{align}
The treatment of the radiation 4-force $G^\mu$ incorporated in our simulations follows the detailed description given by \cite{Sadowski2014,Sadowski2014b} and can be split into
\begin{equation}
    G^\mu = G^\mu_0 + G^\mu_{\rm Compt},
\end{equation}
\noindent where $G^\mu_{\rm Compt}$ is the thermal Comptonization term, while $G^\mu_0$ includes the absorption and scattering opacity terms. For the $G^\mu_0$ term, we assumed a free-free absorption opacity model following Kramer's law
\begin{equation}
\kappa_{\rm a} = 6.4\times 10^{22} \rho T_{\rm g}^{-\nicefrac{7}{2}}\, \mathrm{cm^2 g^{-1}} 
\end{equation}
for a~gas temperature $T_{\rm g}$, and Thomson scattering opacity
\begin{equation}
\kappa_{\rm es}  = 0.34\, \mathrm{cm^2 g^{-1}} \, .
\label{eq:scatop}
\end{equation}

\noindent For the range of densities and temperatures involved, we find that $\kappa_{\rm es} \gg \kappa_{\rm a}$ in the whole domain.

Simulation \Snew{} was performed with an updated version of the \koral{} code, using an improved opacity model described in \citet{Sadowski2017}, including synchrotron and bremsstrahlung (free-free) opacities, as well as energy and angle-averaged Klein-Nishina high-energy correction to the scattering opacity, 
\begin{equation}
\kappa_{\rm es} = 0.34\left[1 + \left(\frac{T_\mathrm{rad}}{4.5 \times 10^8\,\mathrm{K}}\right)\right]^\mathbf{{\textcolor{OliveGreen}{0.86}}}\, \mathrm{cm^2 g^{-1}},
\end{equation}
where $T_\mathrm{rad}$ is the radiation temperature evolved by the code within the $G_\nu$ 4-vector \citep{Sadowski2014}.

\begin{figure*}
    \centering
    \includegraphics[width=\textwidth]{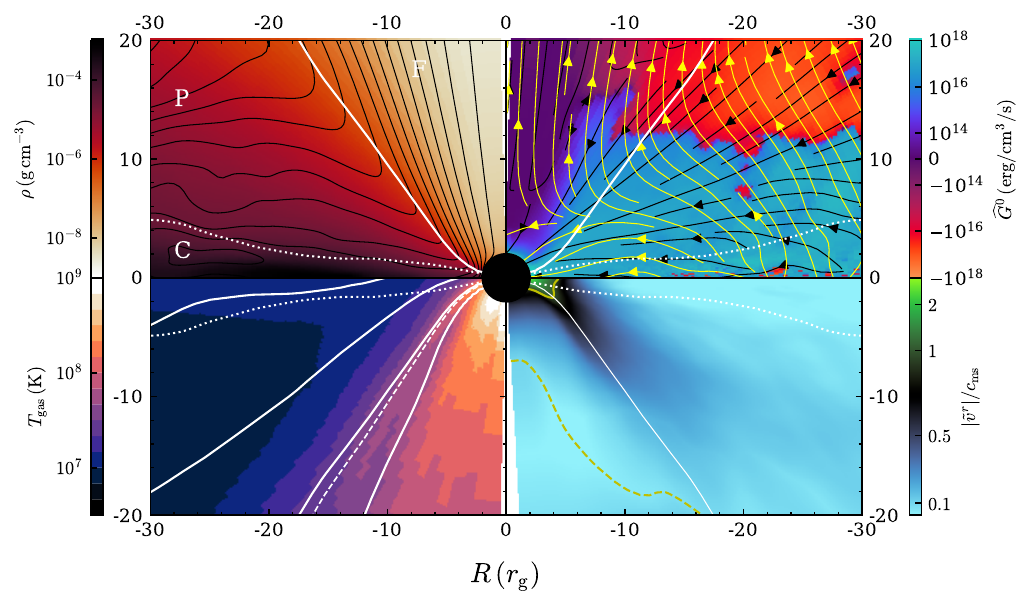}
    \caption{Morphology of puffy disc obtained from \texttt{S06}, averaged over time and azimuth. Two surfaces, corresponding to the density scale height (white dotted line) and the photosphere (white full line), are shown in each panel. They divide the accretion flow into three regions, denoted in the upper-left panel. \textit{Top left:} gas density and magnetic field lines. \textit{Top right:} local dissipation rate $\widehat{G}^0$ and streamlines of gas velocity (black) and radiation flux (yellow). \textit{Bottom left:} gas temperature (discretised) and contours of $\tau=\left(0.1,1,10,100\right)$ from outside the disc towards the equator, and $\tau = \nicefrac{2}{3}$ with white dashed line. \textit{Bottom right:} gas radial velocity in terms of local magnetosonic speed, the yellow contour shows the border where $\nicefrac{|\tilde{v}^r|}{c_\mathrm{ms}}=1$. The yellow dashed line shows the stagnation surface (inflow/outflow boundary), where $\tilde{v}^r$ changes sign. 
    \label{fig:general}}
\end{figure*}

\subsection{Simulation details}

All simulations presented here assume a non-rotating BH of $M=10\,\msol$ (where $\msol$ is the solar mass), using Schwarzschild spacetime.

\subsubsection*{Computational grid}

To cover a large simulation domain with sufficient resolution in the area of interest, we use the Modified Kerr-Schild (MKS) coordinates $\left\{x_t,x_r,x_{\theta},x_{\phi}\right\}$, related to the Boyer-Lindquist coordinates $\left\{t, r,\theta,\phi \right\}$ as
\begin{subequations}
    \begin{align}
        r &= X_0 + e^{x_r} \ ,
        \\
        \theta &= \pi x_{\theta} + \frac{\left(1-H_0\right)\sin\left({2\pi x_{\theta}}\right)}{2}  \,
    \end{align}
\end{subequations}
\noindent where the choice of $X_0$ and $H_0$ concentrates the cells towards the event horizon and the equatorial plane, respectively. 

Since simulations require a significant amount of computational resources, even with a very efficient grid, we model a $\nicefrac{\pi}{2}$ wedge in the azimuthal direction with a periodic boundary condition. \cite{Sadowski2016} showed that this approximation yields results consistent with a full 3D run.

The simulations discussed in this paper were performed with a resolution of ($N_r \times N_{\theta} \times N_{\phi}$) = $320\times320\times32$, $H_0 = 0.875$ and $X_0 = 0$ for simulations \texttt{S06}, \texttt{S09}, and \texttt{S15}, and $384\times 384\times32$, $H_0 = 0.875$ and $X_0 = -1.5$ for simulations \texttt{S04}. In all cases, several cells in the radial direction extend below the event horizon of the central BH.  This grid resolution is sufficient to resolve and sustain the magneto-rotational instability (MRI), with quality factors $Q_\theta > 25$ and $ Q_\phi > 20$ throughout the domain \citep{Balbus1998}.

\subsubsection*{Initial state and simulation duration}

The simulations were initialised with a thick torus \citep{Penna_torus} with the radiation pressure included in its calculations. It is initially located rather far from the BH (with inner edge $r_{\rm in} = 35 \,\rg$) extending to the outer boundary of the domain ($r_{\rm out} = 500\,\rg$), threaded with a~magnetic field of a quadrupole geometry and a~pressure ratio $\nicefrac{\pg}{\pmag} = 20$ in the centre. This state is evolved in an axisymmetric regime for a long time ($\sim 60\,000\,\tg$) to obtain a steadily accreting disc in the innermost region and subsequently extended into 3D. This is then used as the initial state for all simulations discussed in this paper. Since the dynamical timescale for a stellar-mass BH is extremely short relative to the duration of observations ($20\,000\,\tg$ is approximately $1$s for $10\,\rm \msol$ BH), we focus our analysis on time-averaged properties of the disc, and show only time- and azimuth-averaged properties unless stated otherwise. Thus, their duration $\Delta t$ (see \autoref{tab:sims}) does not include the initial run, and the time always labels the beginning of the averaging interval. This setup allows us to study a self-consistently evolved accretion disc in the sub-Eddington regime, supported by MRI-induced turbulence.

Even though the domain is quite large, we are interested only in the innermost initially empty part, where the accreting matter forms the puffy disc. The initial torus serves only as a source of matter to feed the disc, and it is not included in the analysis of the disc properties presented here. The outer torus is regularly restored to its initial state to maintain stable flow into the inner domain.

 We present the results of four simulations, which are summarised in \autoref{tab:sims}.  {We label the simulations \texttt{S04}, \texttt{S06}, \texttt{S09}, and \texttt{S15} after their approximate mass accretion rates ($\mdot \approx 0.45, 0.58, 0.89$ and $1.5$, respectively).} We show time- and azimuth-averaged quantities, and only across the quasi-stable state (up to $r_\mathrm{max}$ for each run) and after an extremely long evolution. Simulation \texttt{S06} was run for $17\,400\, \tg$\footnote{extended by $2\,400\,\tg$ compared to results in \cite{Lancova2019}} long enough to obtain a quasi-steady-state accretion in the inner part $r <r_{\rm max}$. It reaches the normalised mass accretion rate $\mdot$ of 
\begin{equation}
    \mdot= \frac{\dot{M}}{\dot{M}_{\rm Edd}} = \frac{ \dot{M}}{{L}_{\rm Edd} / (\eta \mathrm{c}^2)}  = \frac{0.057 \dot{M} \mathrm{c}^2}{{L}_{\rm Edd}} = 0.58 \ ,
\end{equation}
where $\eta = 0.057$ is an efficiency scaling factor for a non-rotating BH. For comparison, we show results from simulations of \citet{Sadowski2016} with $\mdot = 0.89$ (\texttt{S09}) and a short, slightly super-Eddington simulation with $\mdot=1.51$ (\texttt{S15}). We also included a new simulation for a puffy disc with $\mdot = 0.45$ (\texttt{S04}), which reached a quasi-stable state up to $r_{\max} = 12\,\rg$. Unfortunately, even with the higher resolution and more concentrated grid in the equatorial plane, we were not able to obtain a longer quasi-stable solution for such a low mass accretion rate. Since the total length of this run is much longer than the thermal instability timescale, we include the result in our analysis, especially focusing on the inner edge of the disc.

\section{Properties of the puffy discs}
\label{sec:puffy}

Contrary to analytic modelling, the puffy disc is based on results of 3D global GRRMHD simulations, which include computations of magnetic turbulence. This leads to a solution where the magnetic and radiation pressures strongly dominate over the gas pressure. However, the disc is stable, at the cost of significantly larger vertical extension than expected from analytic models. Furthermore, the properties of the fluid that forms the disc do not change significantly at the ISCO radius. The detailed vertical structure of \texttt{S06} is shown in \autoref{fig:general}.

Throughout the paper, we use capital $R$ for the cylindrical radius, distinguishing it from the Boyer-Lindquist spherical radius $r$. We show all simulation data averaged over azimuth and time.

The most characteristic property of the puffy disc is a distinction between the density height scale $h_\rho$,
\begin{equation}
h_{\rho}\left(R\right) = \sqrt{\frac{\int \rho\left(R,z\right) z^2 {\rm d} z}{\Sigma}},
\label{eq:h_dens}
\end{equation}
where $\Sigma = \int \rho \left(R,z\right) {\rm d} z$ is the vertically integrated density and the height of the photosphere  $h_\tau$, defined through the vertically integrated optical depth calculated against the scattering opacity
\begin{equation}
\tau \left(R,z\right) = \int_{z}^{z_0} \rho \left(R,z'\right) \kappa_{\mathrm{es}} \dif z',
\end{equation}
as a surface where $\tau\left(R,h_\tau\right) = 1$. We chose the outer integration limit as $z_0 = 200\,\rg$, far enough from the equatorial plane to not affect the resulting value of the photospheric height.

This distinction is absent in vertically averaged analytic disc models, where these two surfaces are assumed to coincide. The two surfaces divide the structure of a puffy accretion disc into three zones indicated in the top-left panel of \autoref{fig:general}: a geometrically thin, optically thick disc core (C), a geometrically and optically thick puffy region (P), and an optically thin funnel region (F). This structure is somewhat similar to the predictions of \citet{Zhang2000} based on X-ray spectral modelling of BH XRBs. 

\begin{figure}
    \centering
    \includegraphics[width=\columnwidth]{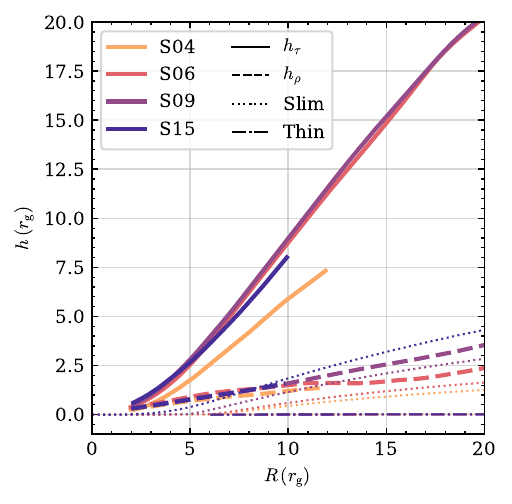}
    \caption{Vertical thickness of the accretion discs. Full solid lines correspond to the thickness of the photosphere (\texttt{SO6} and \texttt{S09} lines nearly coincide), and the thick dashed line to the scale height from simulated discs (the lines of all four models nearly coincide), with mass accretion rate denoted by the line colours. The dotted thin lines (increasing with $\dot m$) show the thickness of the slim disc, and the dash-dotted line the thin disc. The thin and slim disc model curves were calculated for a non-rotating BH with mass $M = 10\,\msol$, viscous $\alpha = 0.1$, and mass accretion rate the same as the corresponding simulation, denoted by the same colours. The simulated data are shown up to the radius where the quasi-stable state was reached.}
    \label{fig:Hbyr}
\end{figure}

\subsection{The disc's inner edge}
\label{sec:inneredge}

The inner edge of the standard thin disc is assumed to be at the ISCO, with the condition that the stress vanishes there, and this serves as the boundary condition for solving the disc equation. In the slim disc model, the inner edge is found self-consistently as the location of the sonic point, and it is sensitive not only to the value of $\mdot$, but also to the assumed viscous $\alpha$. For some combinations of these parameters, the inner edge may even be located at radii larger than ISCO.

Simulation results show that the puffy discs are optically thick up to the event horizon, making it difficult to define the inner edge of the thermal emission. The local thermal equilibrium, required to produce a black-body spectrum, holds if the matter's radial velocity is subsonic. We can also find the radius at which the gas temperature starts to vary significantly from the analytic disc models and use that value as the inner edge of the thermal emission. In the equatorial plane, this radius is around $10\,\rg$ (see the top panel of \autoref{fig:T_Sigma}), where, in analytic models, the temperature reaches a maximum, but it continues to increase in the simulations. Furthermore, in the case of the  magnetised and vertically stratified puffy disc, we need to locate a magnetosonic surface where the gas radial velocity equals the quadrature sum of the fast Alfv\'{e}n ($c_{\rm A}$) and sound ($c_{\rm s}$) speed, giving the magnetosonic speed $c_{\rm ms}$ \citep{Gammie2003},

\begin{align}
    c_\mathrm{s}^2 &= \frac{\gamma \pg}{\rho + u_\mathrm{int} + \pg}, \\
    c_{\mathrm{A}}^2 &= \frac{b^2}{b^2 +\uint + \rho + \pth},\\
   c_\mathrm{ms} &= \sqrt{c_\mathrm{s}^2 + c_{\mathrm{A}}^2},
\end{align}
\noindent  {where the variables were introduced in \autoref{sec:koral}.}

\begin{figure}
    \centering
    \includegraphics[width=\linewidth]{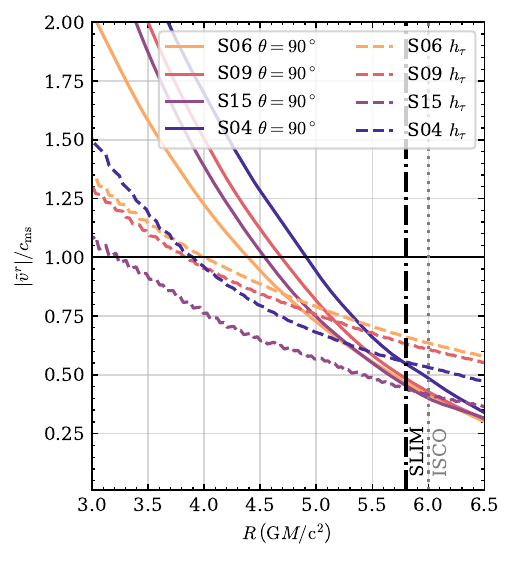}
    \caption{The radial velocity to the local magnetosonic speed ratio in the innermost parts of the accretion disc for all simulations. The full lines show the values at the equator, while the dashed lines show the values on the photosphere. The black dot-dashed vertical line is the sonic point location of the slim disc model (which coincides for all simulated values of mass accretion rate).  The dip in the \texttt{S15} data is caused by the short duration of the simulation.}
    \label{fig:cms}
\end{figure}

The ratio of the gas radial velocity $\tilde{v}^r$ and the magnetosonic speed $c_\mathrm{ms}$ is shown in the bottom-right panel of \autoref{fig:general}. \autoref{fig:cms} then shows the values of the $\nicefrac{|\tilde{v}^r|}{c_\mathrm{ms}}$ ratio on the equator and on the photosphere close to the ISCO radius and the exact location of the sonic point on the equator and the photosphere for all simulations. The location of the slim disc sonic points (which coincide for the values of used $\mdot$) is compared. The locations of the photospheric sonic points in the simulations, $R_{\rm sp}$, are also summarised in~\autoref{tab:sims}. 
Thus, the inner edge of a puffy disc, defined as a magnetosonic radius, is located significantly closer to the BH's event horizon than the inner edge in the analytic models.

The location of the inner edge is essential for the observable properties of the disc and for modelling the luminous state of X-ray binaries. As we demonstrated in \citet{Wielgus2022}, the standard tools for fitting the X-ray source parameters are inadequate when applied to puffy disc simulations, particularly for spin estimation. Since the ISCO location depends exclusively on the metric of the central BH, finding its position through spectral fits to the standard disc model appears to be an excellent method of estimating the spin. However, the results of both numerical simulations and analytic modelling show that in the luminous regime, the inner edge of a disc does not lie on the ISCO, and thus this method may be inaccurate \citep{2002ApJ...573..754K,2024MNRAS.531..366M}.

\subsection{Disc thickness and density}

The geometrical thickness of the puffy disc photosphere arises from the magnetic field, which provides an additional pressure component in the midplane of the disc, thereby stabilising it against thermal instabilities. To generate enough pressure, the magnetic field must have a strong vertical or radial net flux, which is not possible with a dipolar topology, as was shown in \citet{Sadowski2016,Lancova2019} and also \cite{2022ApJ...939...31M}. We adopted this solution in our simulations.

The disc core extends relatively high: for \texttt{S06}, $h_\rho / R \approx 0.10$,  and contains $\sim$ 80\% of the total column-integrated mass of the disc, with the maximum density in the equatorial plane of $\sim 10^{-3}$\,g\,cm$^{-3}$. Compared to the vertical structure of the slim disc model, the core region is more elevated at the same $\mdot$. \autoref{fig:Hbyr} compares the simulated disc's vertical extension with the slim and thin disc profiles. All simulations show larger photospheric heights than those of the analytic models, and they are almost independent of $\mdot$. 

The bottom panel of \autoref{fig:T_Sigma} shows the vertically integrated density comparison with the analytic models. The slim disc density drops around the ISCO, and in the limit of low $\mdot$ it approaches the thin disc, sharply terminating at the ISCO. Numerical models show significantly lower surface density than expected in analytic models, indicating a higher accretion velocity (see top panel of \autoref{fig:velprofile}), higher effective viscosity coefficient $\alpha$ (see \autoref{sec:alpha}), magnetic pressure support, and strong advection, all valid for puffy disc simulations.

The photospheric height gives the observable surface $h_\tau / R \approx 1$ for all simulations, and it defines the puffy region, located between $h_\tau$ and $h_\rho$. \texttt{S04} shows $h_\tau$ lower by about 30\% than the other models, which hints at a transition to a thinner disc for even lower mass accretion rates -- we aim to investigate this in future simulations.

Although the simulated disc is substantially geometrically thicker than expected from standard analytic models, the presence of the puffy region is qualitatively consistent with the calculations of stationary stratified discs of BH XRBs \citep[][]{Gronki2020}, where a similar layer is referred to as a warm corona. \citet{2024ApJ...962..101Z} recently demonstrated that spin measurement using a similar topology with a warm corona gives spin values significantly lower than those obtained using other methods. It agrees well with the puffy disc results for spin measurement \citep{Wielgus2022,2023AN....34430023L}.

\subsection{Radiation, pressures, temperature {, and opacity}}

\begin{figure}
    \centering
    \includegraphics[width=1\linewidth]{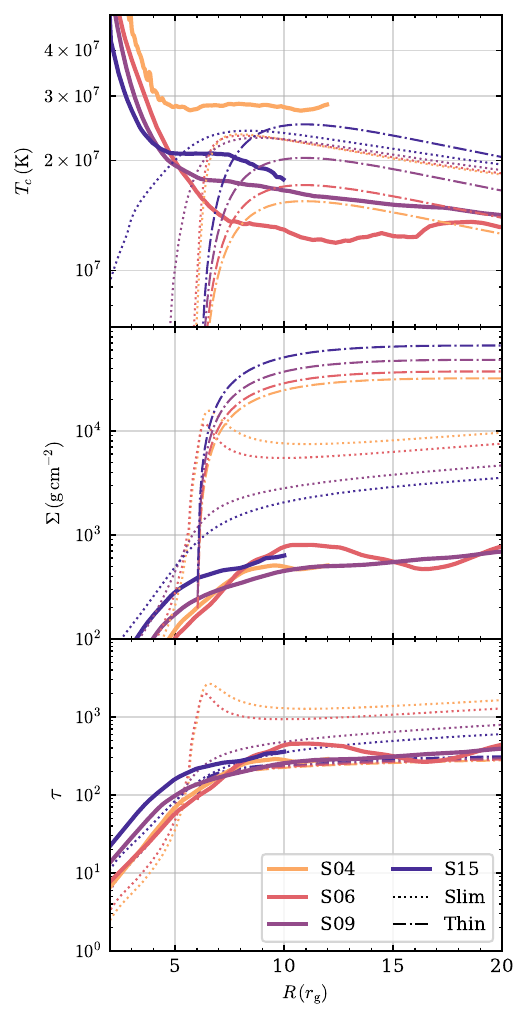}
    \caption{\textit{Top:} Central gas temperature $T_c$ in simulations (full lines), slim disc (dotted lines), and thin disc (dot-dashed lines).
    \textit{Middle:} Vertically integrated (surface) density $\Sigma$ from simulations (full lines), slim (dotted lines), and thin (dash-dotted lines) disc. 
     {\textit{Bottom:} The optical depth at the equator compared to slim and thin disc models. For the thin disc, $\tau$ is obtained directly from the disc structure solution; for the slim disc, it is estimated as $\tau = \nicefrac{1}{2}\,\kappa_{\mathrm{es}}\Sigma$ from the surface density}. The parameters for the thin disc models are the same as in \autoref{fig:Hbyr}.}
    \label{fig:T_Sigma}
\end{figure}

In the optically thick regions, the radiation approximately follows the negative of the density gradient, emerging in the funnel region, where it is collimated along the axis. The upper right panel of \autoref{fig:general} shows the local dissipation rate, which can be approximated by the time component of $G^\mu$, see eq. (\ref{eq:consv}). In the dense disc region around the equatorial plane, the gas energy is dissipated into radiation, and only in the upper regions does the radiation heat the gas. In the hot, low-density funnel region, the gas and radiation are not strongly coupled.

The bolometric luminosity from simulations may be estimated as an integrated radial radiation flux in the optically thin region at a certain radius, for which we choose $r_{\mathrm{lum}}=25\,\rg$ to reduce the influence of the outer torus,  
\begin{equation}
    L(r_{\mathrm{lum}}) =2{\pi}\int_{0}^{\pi}\sqrt{-g}\,\left(-R^{r}{}_{t}\right)_{(\rm \tau < 1)}\dif\theta,
\end{equation}
and the values for each simulation are summarised in \autoref{tab:sims}. However, full radiation transfer post-processing is needed for a proper luminosity estimation, and furthermore, the M1 closure used in \koral{} is known to overstate the radiation energy density in the funnel \citep{2020ApJ...901...96A}. 

 {We choose $r_\mathrm{lum} = 25\, \rg$, well inside the inner edge of the outer torus (at $r_\mathrm{in} = 35\, \rg$), to minimise its contribution to the estimated luminosity, but far from the region where the radiation is advected into the BH. However, this is a lower limit and a rather crude estimate. An accurate luminosity estimate requires full radiation transfer post-processing, as discussed in \citet{Wielgus2022}.}

\begin{figure*}
    \centering
    \includegraphics[width=\textwidth]{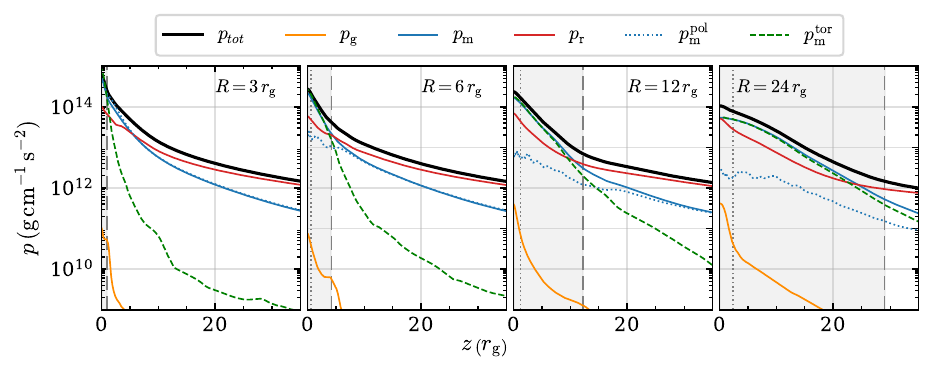}
    \caption{Vertical distribution of pressure and its components on four cylindrical radii in \texttt{S06}. The dashed grey line labels the location of the photosphere, and the dotted line is the density scale height. Grey-shaded areas are located under the photosphere.}
    \label{fig:vertical_pressure}
\end{figure*}

The upper panel of \autoref{fig:T_Sigma} shows the central (equatorial plane) gas temperature $T_{\mathrm{c}}$ for all simulations and its comparison to analytic disc models. It is apparent that, in contrast to the analytic models, the temperature does not drop in the inner region, which is mainly attributed to the fact that the simulated discs are optically thick up to the BH horizon. Outside the ISCO, the simulations show temperatures comparable to those expected for thin or slim discs, except for \texttt{S04}, where the temperature is significantly higher. These temperatures are a few times higher than the prediction of \cite{Zhang2000}, which can be related to the fact that we only probe the innermost part of the accretion flow, whereas \cite{Zhang2000} estimates the observed temperature from a global radial structure of a standard thin disc.  {In the bottom panel of \autoref{fig:T_Sigma} we compare the optical depth $\tau$ at the equator of the puffy disc simulation to the analytical models. The simulations agree well with the thin disc model but less so with the slim disc; note that the slim disc optical depth is estimated as $\nicefrac{1}{2}\,\kappa_{\mathrm{es}}\Sigma$ from the surface density, whereas for the thin disc $\tau$ it
is a direct output of the disc structure solution.}

The vertical temperature distribution in \texttt{S06} is shown in the bottom left panel of \autoref{fig:general}, demonstrating that there exists a layer of hot gas on top of the disc, close to the photosphere, in regions with $\tau < 10$. This again resembles the warm corona model, and agrees with vertical structure calculations of a stratified magnetically heated disc with atmosphere cooled by Comptonization \citep{Gronki2020}. The optical depths of the warm corona base, obtained by those authors, were $\tau \sim 10$ for accretion rates $\dot m \geq 0.3$ and for relatively high magnetisation.

We investigate the detailed vertical distribution of pressures in \texttt{S06} in \autoref{fig:vertical_pressure}. Near the photosphere, radiation pressure begins to dominate over magnetic pressure. Gas pressure is negligible throughout the domain, $\pr \gg \pg$. In the magnetic pressure, we separated the poloidal and toroidal contributions,
\begin{align}
    p_{\rm m}^{\rm pol} &= \left(b_{r}b^{r} + b_{\theta}b^{\theta}\right)/2,\\
    p_{\rm m}^{\rm tor} &= b_{\phi}b^{\phi}/2, 
\end{align}
and we can see that the poloidal component dominates in the funnel region, while in the disc itself, the toroidal component is prominent.
The transition corresponds to the approximate location of the photosphere.

In the simulated disc, the magnetic pressure $\pmag$ dominates over the thermal pressure $\pth = \pr + \pg$ by a factor of about 2 in the optically thick region. The height-integrated magnetic and radiation pressures are comparable, which agrees with the stabilisation criterion of \citet{Sadowski2016,Sadowski2016b}. We summarise the averaged density-weighted $\beta = \pth/\pmag$ and $\beta_{\rm rad}(=\pr / \pg)$ in the simulations in \autoref{tab:sims}.

Although magnetic pressure dominates in the disc and the magnetisation is high, the disc does not reach the magnetically arrested \citep[MAD;][]{2003PASJ...55L..69N} state, most likely due to the magnetic field quadrupole topology. During simulation \texttt{S06}, the average value of the magnetic flux through the BH horizon does not reach more than 15\% of the MAD limit \citep{2011MNRAS.418L..79T}, with a maximum reaching 25\%. A strong poloidal component is needed to stop accretion in the MAD state, but in the puffy disc, the toroidal component dominates (see \autoref{fig:vertical_pressure}). Due to the significant alteration in the magnetic field direction around the photosphere (see top-left panel of \autoref{fig:general}), a current sheet may lead to magnetic reconnection events and particle acceleration at this location. However, in our simulations, the use of ideal MHD approximation and the resolution do not allow us to explore this phenomenon thoroughly \citep{2022ApJ...933...55C,2024MNRAS.52710151K}.

\subsection{Accretion flow}
\label{sec:gasspeed}

\begin{figure}
    \centering
    \includegraphics[width=\columnwidth]{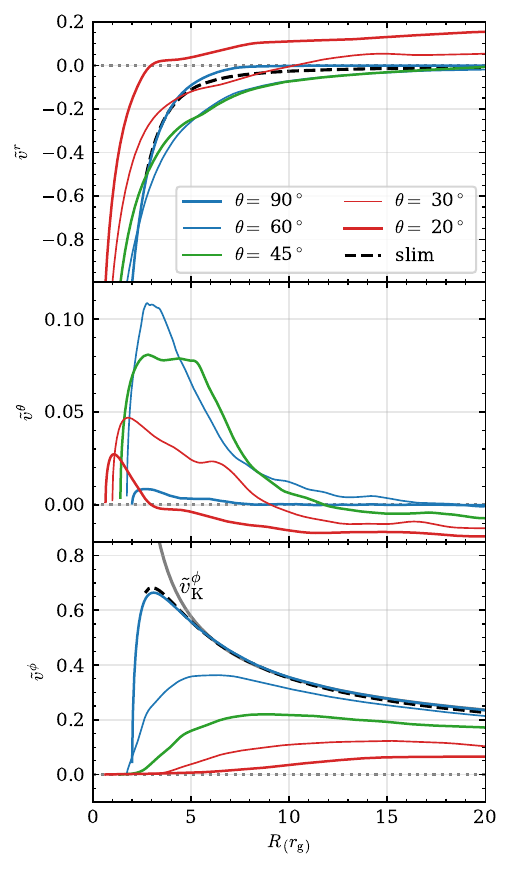}
    \caption{Velocity profiles from \texttt{S06} along different inclinations  { from close to the axis ($\theta =20^\circ$) to the equator ($\theta =90^\circ$)} as a function of the cylindrical radius. The black-dashed line in the first and last panel corresponds to the slim disc solution with the same mass accretion rate. Positive radial velocity $\tilde{v}^r$ indicates outflows. Keplerian profile is denoted by $\tilde{v}^\phi_{\rm K}$ in the last panel.}
    \label{fig:velprofile}
\end{figure}

We compare the components of gas velocity profiles from \texttt{S06} for five inclinations from the equatorial plane ($\theta = 90^\circ$) to the funnel ($\theta=20^\circ$), with highlighted values around the photosphere ($\theta = 45^\circ$), to those of analytic models of the same $\dot{m}$ in \autoref{fig:velprofile}. The gas velocity is also shown in the upper-right panel of \autoref{fig:general} with black streamlines. We transform the 4-velocity components from simulations to the physical velocity components measured in the frame of the zero angular momentum observer
\begin{equation}
    \tilde{v}^i = \left(-\frac{g_{ii}}{g_{tt}}\right)^{\nicefrac{1}{2}}\frac{u^i}{u^t} .
\end{equation}

The slim disc model is based on a solution of 1D equations of motion, so the only non-azimuthal component of velocity is the radial one, and it is independent of $z$. Inclusion of radial pressure gradients allows the solution to be extended to the plunging region inside the inner edge of the disc. The standard thin disc solution is likewise a 1D model, however it has an unrealistic value of radial velocity at the inner edge, an artefact of the no-torque boundary condition at the ISCO, and for this reason has not been extended to the plunging region. Thus, we can only compare the puffy disc velocity field with the azimuthal and radial velocity of the slim disc, for which the components depend only on $R$ and not on $\theta$. 

For small $h_\rho/R$ the disc must be supported radially by rotation, with the azimuthal velocity coinciding with the Keplerian one to a high degree of accuracy \citep{Shakura1973,KK2000,2000astro.ph..4129P}. This applies equally to the thin disc model, the slim disc, and the core of the puffy disc. Once that rotational support is removed (near the ISCO, the last stable orbit of a test particle), the fluid must plunge in towards the BH. This is apparent in the upper panel of \autoref{fig:velprofile}, where good agreement is observed between the velocity of the slim disc and the equatorial velocity of the puffy disc. The agreement extends to the azimuthal velocity as well (bottom panel). Both discs indicate an azimuthal velocity profile in the equatorial plane close to the Keplerian one above the ISCO but slowing towards zero near the BH horizon.

In simulations, the radial infall of the gas at the equator starts to be non-negligible ($\tilde{v}^r < - 0.1$) at around $r\sim8\,\rg$ rather than under the ISCO radius. However, the observable surface is located at about $\theta \sim 45^\circ$, where the radial velocity is high already at $r \sim 20\,r_g$. Under the ISCO, the equatorial radial velocity agrees remarkably well with the slim disc model. 

Recently, a trans-ISCO model of the accretion disc was developed by \citet{2023MNRAS.521.2439M}, characterising the stress on the ISCO by parameter $\delta_\mathcal{J}$, which corresponds to the angular momentum loss between the ISCO and the BH horizon at $r_{\mathrm{h}}$. The ISCO stress parameter is defined as
\begin{equation}
\delta_\mathcal{J} = \frac{\langle u_\phi\rangle_\rho(r_{\mathrm{h}}) - \langle u_\phi\rangle_\rho(r_{\mathrm{ISCO}})}{\langle u_\phi\rangle_\rho(r_{\mathrm{ISCO}})},
\end{equation}
and the puffy disc values are summarised in \autoref{tab:sims}. For \texttt{S06}, $\delta_\mathcal{J} = 0.16$, which is much higher compared to the results of non-radiative GRMHD simulations with cooling function, where $\delta_\mathcal{J} \sim 0.002 - 0.05$ \citep[e. g.,][]{2008ApJ...687L..25S,2025arXiv250513701R}. This value affects the modelled emission from the plunging region, which can be used to fit the observed data \citep{2024MNRAS.531..366M,2024MNRAS.533L..83M}.

\subsection{Advection and photon trapping}
\label{sec:adv}

The advection of matter plays a significant role in the puffy disc, even at larger radii than predicted by the slim disc model, given the greater vertical thickness. The top-right panel of \autoref{fig:general} shows the streamlines of both gas velocity (black) and radiation flux (yellow). In the region where photons are trapped and advected with the fluid, the radiation flows in the same direction as the gas. In  \texttt{S06}, photons are trapped in the disc up to $R \sim 20\,\rg$. However, advection alone is not sufficient to stabilise the disc -- a simulation with the same configuration but a dipolar magnetic field in \cite{Sadowski2016} did not yield a stable solution. 
Furthermore, it was demonstrated that a radial net flux of the magnetic field is necessary to stabilise the flow \citep[][]{2022ApJ...939...31M}. In the funnel region above the photosphere, about $z\sim 7\,\rg$, there is a gas stagnation surface, denoted by a yellow dashed line in the bottom-right panel of \autoref{fig:general}, where the radial component of the gas velocity changes sign. Apart from the compact near-equatorial photon trapping region, radiation escapes to infinity. This includes the optically thick core and the puffy regions.

\begin{table*}
\begin{threeparttable}
    \caption{Overview of the puffy disc models: mass accretion rate $\mdot$, time-averaging window $\Delta t\,$, the maximal radius where a quasi-stable state was reached $r_{\mathrm{max}}$, thickness of the core $h_\rho/R$, and the photosphere $h_\tau/R$, bolometric luminosity $L/L_{\mathrm{Edd}}$ at $r_{\mathrm{lum}} = 25\,\rg$, the gas temperature at the center $T_{\mathrm{c}}$, magnetic $\beta = \pth/\pmag$, radiation $\beta_{\mathrm{rad}} = \pr/\pg$, radius of the magnetosonic ``point'' (cylinder) at the photosphere $R_{\mathrm{sp}}$, the density-weighted average of viscous $\alpha$ between the sonic point and $r_{\mathrm{max}}$, and the ISCO stress parameter $\delta_\mathcal{J}$. If not stated otherwise, those are shown for the quasi-stable region of the disc.\label{tab:sims}}
\begin{tabular}{c c c c c c c c c c c c c}
\hline
Simulation & $\mdot$ & $\Delta t\, (\tg)$ & $r_{\max}\,(\rg)$& $h_\rho/R$ & $h_\tau/R$ & $L/L_{\mathrm{Edd}}$ & $T_{\rm c}$ ($10^7$ K) & $\beta$ & $\beta_{\rm rad}$ & $R_{\rm sp}\,(\rg)$ & $\langle \alpha \rangle $ & $\delta_\mathcal{J}$\\\hline \hline
\texttt{S04} & 0.45 & 7000  & 12           & 0.13   & 0.7    & 0.23 & 3.29 & 0.48 &   390  & 3.90   & 0.26        & 0.10\\ \hline                
\texttt{S06} & 0.58 & 17400 & 30\tnote{1}  & 0.16   & 1.1    & 0.36 & 1.52 & 0.42 &   679  & 3.99   & 0.19        & 0.16\\ \hline                
\texttt{S09} & 0.89 & 20000 & 30\tnote{1}  & 0.16   & 1.1    & 0.35 & 1.70 & 0.51 &   702  & 3.84   & 0.20        & 0.12 \\ \hline
\texttt{S15} & 1.51 & 5600  & 10           & 0.15   & 0.9    & 0.44 & 2.27 & 0.40 &  1073  & 3.15 & --\tnote{2} & 0.10\\ \hline
\end{tabular}
    \begin{tablenotes}
        \item[1] the maximal radius before the outer torus influences the results.
        \item[2] run \texttt{S15} is not long enough to measure $\alpha$ reliably.
    \end{tablenotes}
\end{threeparttable}
\end{table*}

\section{Effective viscosity}
\label{sec:alpha}

The primary mechanism of angular momentum transport in accretion discs is now known to be the MRI-driven turbulent viscosity \citep{Balbus1991,Balbus1998}. Already before the role of the MRI was recognised, viscosity has been parameterised with the $\alpha$ parameter in \citet{Shakura1973}. There have been conflicting reports of its precise value, with simulations preferring $\alpha \approx 0.01$ and observations reporting $\alpha \approx 0.1$ \citep{King2007}.

To estimate $\alpha$ from our simulation, we followed the approach outlined in \cite{Penna2013}, which also allows us to separate the contribution to $\alpha$ from Reynolds and Maxwell stresses. We can split the stress-energy tensor into its corresponding components,
\begin{align}
&T_{\mu\nu}^{\rm {Rey}} = \left(\rho + \uint + \pg\right)u_\mu u_\nu + \pg g_{\mu\nu},\\
&T_{\mu\nu}^{\rm {Max}} = b^2 u_\mu u_\nu - b_\mu b_\nu + \frac{1}{2}b^2g_{\mu\nu}.
\label{eqn:ReyMax}
\end{align}

\begin{figure}
    \centering
    \includegraphics[width=\columnwidth]{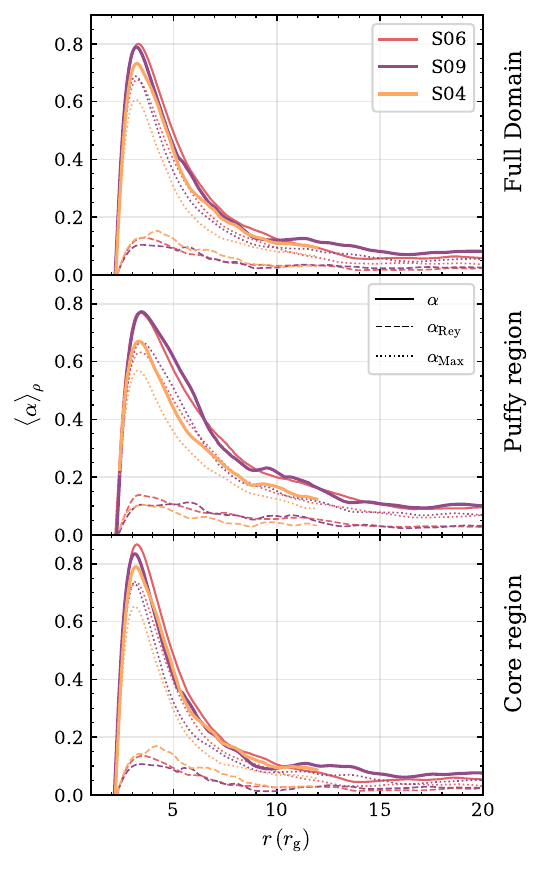}
    \caption{Density-weighted shell average viscous $\alpha$ radial profile for total (full lines), Reynolds (dashed lines) and Maxwell (dotted lines) $\alpha$ in the three regions of the puffy disc. Values from different simulations are labelled by colours.}
    \label{fig:alpha}
\end{figure}

\noindent First, we define the frame of a mean flow (denoted by hats), using the gas four-velocity averaged from the full duration of the simulations, and tetrad $e^\mu_{\hat{\alpha}}$ \citep[][Appendix B]{Kulkarni2011}. In this frame, we can recognise both the magnetic stresses and the stresses arising from the gas turbulence, which would be zero in an instantaneous fluid frame in each individual snapshot. Then, 100 evenly spaced 3D snapshots from the run are selected, and the stress-energy tensor components, $\alpha$, $\alpha_{\rm Rey}$, and $\alpha_{\rm Max}$ are evaluated in each of them, as
\begin{equation}
    \alpha = \dfrac{T_{\hat{r}\hat{\phi}}}{p_\mathrm{tot}},
\end{equation}
\begin{align}
    &\alpha_{\rm Rey} = \dfrac{T_{\hat{r}\hat{\phi}}^{\rm {Rey}}}{p_{\mathrm{tot}}},\quad \alpha_{\rm Max} = \dfrac{T_{\hat{r}\hat{\phi}}^{\rm {Max}}}{p_{\mathrm{tot}}},
    \label{eqn:alpha}
\end{align}
\noindent using the total pressure, $p_{\mathrm{tot}} = \pr + \pg + \pmag$. Subsequently, these 3D values are averaged over time and azimuth, and a density-weighted shell average gives the radial profile shown in \autoref{fig:alpha}.

We compare the values of total, Reynolds, and Maxwell $\alpha$ in the different regions of the puffy disc in \autoref{fig:alpha}, where they are integrated in the whole domain, only in the core, or only in the puffy region. We show that the Maxwell stress is the main source of  viscosity in the simulations and that far from the BH, $\alpha$ in the disc is close to the observed value of $\alpha \sim 0.1$, mainly due to the contribution of $\alpha_{\rm Max}$, which is important in both the core and the puffy region. We also see that the viscosity in the disc core is lower than in the puffy region, outside the plunging region, which shows that the elevated low-density region contributes significantly to the total viscosity.

In agreement with the other simulation-derived radial profiles of $\alpha$ \citep[e.g., ][]{Penna2013,2008ApJ...687L..25S,Jiang2019,2025MNRAS.542..377R}, we see a significant increase in viscosity in the innermost regions, which is in stark disagreement with the analytic models, where $\alpha={\rm const}$ everywhere by assumption. The peak is primarily due to an increase in Maxwell stress and also affects regions inside the ISCO (or the puffy disc inner edge).
We discuss this behaviour and introduce a universal $\alpha$ formula in \cite{2026arXiv260310997A}.

\section{Discussion and Conclusions}
\label{sec:conclusions}

The puffy disc simulations were performed using a 3D radiative GRRMHD code \koral{} in a regime where analytic models are not applicable. They were initialised with a large torus far from the central BH, and thus show the self-consistently developed accretion flow. The radiation pressure strongly dominates over the gas pressure throughout the domain; however, the magnetic pressure is of the same order of magnitude as the radiation pressure and significantly alters the flow structure while stabilising it. Most notably, the puffy disc photosphere is geometrically thick with $h_\tau/R \sim 1$ almost independently of the mass accretion rate. This feature has significant implications for the observable picture of a system with a puffy disc, as discussed extensively in \citet{Wielgus2022}. In this work, we compared the puffy disc with analytic and semi-analytic models of a thin and slim disc to demonstrate that neither can adequately replace the complex results of numerical simulations.

The puffy disc properties strongly deviate from analytic models, even when a vertical structure is included in the latter case; however, its mass accretion rate falls within the range where both models are commonly used to interpret observations. Most notably, the extended vertical structure, significant accretion velocity, and an inner edge located close to the BH, well inside the ISCO, strongly disagree with the analytic models. These properties are partially linked to the strong magnetisation of the accreting material, which, at the same time, serves as the stabilisation mechanism. Notably, these properties vary only slightly in the simulated $\mdot$ range. We acknowledge that a broader parameter space needs to be explored; however, the computational costs of such long 3D radiative simulations are exceptionally high.

A significant property of the puffy disc is that the inner edge, defined as the magnetosonic point location, is located well inside the ISCO in contrast to the standard thin disc assumption that the stress vanishes and the disc is truncated at the ISCO. Overall, we did not find any significant change in the accreting fluid properties at the ISCO radius. As discussed by \citet{2002ApJ...573..754K} or \citet{2010A&A...521A..15A}, the accretion disc inner edge may be defined in several ways. Here, we discuss these different definitions and where they may be found in the puffy disc:

\begin{itemize}
    \item The turbulence edge, defined as the boundary between the region where magnetic field evolution is driven by the turbulence, and one where it results from flux freezing. This boundary corresponds to a region where  $\alpha_{\rm{Max}}$ starts to deviate from its constant turbulence-induced value; in the puffy disc simulations, this occurs at $r_{\rm{in}}^{\rm{tur}} \sim 10 \,\rg$ independently of $\mdot$ (see the upper panel of \autoref{fig:alpha}). 
    \item The stress edge can correspond to the location of the magnetosonic point $R_\mathrm{sp}$, if the flow is axially symmetric and steady -- which is mostly satisfied in the puffy disc simulation. The magnetosonic point is located deep inside the ISCO radius for all simulations, as shown in \autoref{fig:cms}, independently of the azimuthal location, whether in the equatorial plane or on the observable surface. The values are summarised in Table~\ref{tab:sims}.
    \item The reflection edge located on the radius where the disc becomes optically thin to electron scattering---which we do not observe in the puffy disc simulations, since $\tau > 1$ all the way to the inner boundary located at the BH event horizon at $2\,\rg$. The fact that the disc is optically thick everywhere may lead to its misinterpretation as a signature of a small ISCO radius of a highly spinning BH  when using spectral-fitting methods for spin estimation.
    \item The radiation edge is the innermost radial distance from which photons produced in the disc can escape to infinity.  In the puffy disc simulation, the advection of radiation is very strong, and the produced photons are trapped in the fluid and carried inward, which is shown in the upper-right panel of \autoref{fig:general}. This panel also shows that the accretion energy is dissipated into radiation throughout the entire disc. Thus, in the puffy disc, the radiation is produced up to the inner boundary, but in the innermost region (a few $r_g$ from the BH), most of it falls into the BH and disappears below the horizon. At the same time, some of the radiation produced at larger radii is advected and shifted inward, potentially leading to a discrepancy between the observed thermal spectrum and the one expected from analytic models. The ray-traced images of the puffy disc presented in \citet{Wielgus2022} also show that the emissions seem to be coming from deep inside the ISCO radius.
\end{itemize}

The puffy disc simulations also revealed that another of the analytic model assumptions is incorrect -- the value of the viscous $\alpha$ is not constant throughout the disc. We calculated the $\alpha$ value from time- and azimuth-averaged data from the self-consistent simulations. In agreement with various other works, we found that $\alpha$ rises steeply in the innermost region, primarily due to the increase in magnetic-field-driven Maxwell stresses. 
This rise of magnetic stresses with radius is expected, since the flow is strongly magnetised and is thinning towards the BH horizon (see the discussion in \citealt{2025arXiv250513119M}). We plan to explore the implications of this $\alpha$ parameter variation for  analytic models and the observational consequences in future works, following \citet{2026arXiv260310997A}.

In contrast to the standard thin and slim disc analytic models, the puffy disc is a three-dimensional model, and thus also shows non-zero vertical velocity, which is particularly strong near the photosphere (see the middle panel of \autoref{fig:velprofile}). This strong inflow, more or less parallel to the observable surface, together with the geometrical thickness, may help explain some peculiar properties of the high polarisation degree observed in the XRBs in the soft state in the X-ray domain \citep{2024A&A...688L..27V,2024ApJ...969L..30S,2024ApJ...964...77R}. \citet{2023ApJ...957....9W} argued that the disc thickness itself is not sufficient to explain the observed high polarisation degree; however, they used a canonical slim disc model with rather low thickness ($h < 0.3 \,R$) compared to the puffy disc simulations.

\begin{figure}
    \centering
    \includegraphics[width=1\linewidth]{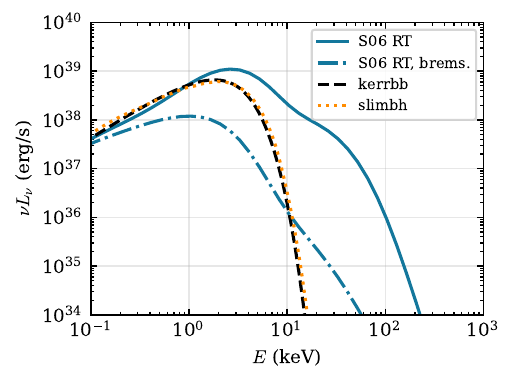}
    \caption{Comparison of the observable spectral energy distributions of the analytic disc models and a puffy disc numerical simulation. We assumed a $10\,\msol$ BH with $\dot{m}=0.6$ viewed at the inclination of $i=10^\circ$. The models are the $\alpha$-independent \texttt{kerrbb} thin disc model (black dashed line), and the \texttt{slimbh} spectrum (dotted orange line). The continuous blue line shows the total spectrum from post-processed \texttt{S06}, including the contribution from comptonized radiation. The bremsstrahlung component from \texttt{S06}, associated with the thermal emission from the disc, is shown by the blue dash-dotted line.
    \label{fig:spectra}}
\end{figure}

The discrepancies between a puffy disc and analytical models affect the observable spectral properties. In \citet{Wielgus2022}, we examined this in detail, however, here we specifically illustrate the differences between the thermal spectral components. In \autoref{fig:spectra}, we show the $\alpha$-independent spectrum from a thin disc \citep[\texttt{kerrbb;}][]{Li2005}, and from a slim disc \citep[\texttt{slimbh;}][]{Straub2011}. We assume the same BH mass and mass accretion rate ($\mdot = 0.6$) as in the simulation, and a default spectral hardening factor $f_\mathrm{col} = 1.7$ \citep{1995ApJ...445..780S}. In the sub-Eddington regime, thin and slim disc spectra almost coincide and the latter only weakly depend on the effective viscosity (we assumed $\alpha = 0.1$ for the plot). Additionally, the figure shows the spectrum obtained from full radiative transfer calculations using the \texttt{HEROIC} code \citep{Narayan2016} for the puffy disc simulation \texttt{S06}. We show the total spectrum (labelled \texttt{S06 RT}), which includes the inverse Compton component, and also the bremsstrahlung component computed without comptonization, as a proxy of the thermal disc emission (labelled \texttt{S06} RT brems.).

\autoref{fig:spectra} clearly shows that even without accounting for Comptonization, spectra from a puffy disc simulation are significantly different from predictions of the analytic disc models. The puffy disc spectrum develops a high-energy tail that, in the absence of the inverse-Compton effect, is attributed to the hot optically thick flow in the plunging region, see the central temperatures shown in \autoref{fig:T_Sigma} as well as the photospheric temperatures in Figure 2 of \citet{Wielgus2022}. At the same time, the peak of the spectral energy distribution flattens with the increased advection limiting the radiative efficiency in the simulated discs. It remains to be shown in future works whether a slim disc model, calculated assuming a non-constant radial profile of viscosity $\alpha$ and/or including the magnetic pressure component, could better reproduce the properties of advanced numerical models and thus serve as their effective approximation.

Simulations of puffy discs demonstrated that advanced numerical models of the soft luminous state in XRBs differ significantly from standard analytic models. Thus, it is not surprising that the relation between observables and model parameters, such as the BH spin, differs as well. The simulations, although their duration appears extremely short in physical units, show a stable behaviour of matter under extreme conditions of a radiation-pressure-dominated luminous disc, which is unstable when modelled analytically. The stabilising mechanism is the vertical net flux of the magnetic field, which, at the same time, causes a high vertical extension of the optically thick flow, a stratified vertical structure, and large inflow velocities close to the photosphere as well as conical outflow near the axis.

 The observed luminosity of a puffy disc depends strongly on the viewing angle, as can be clearly seen in Figure 5 of \cite{Lancova2019} and Figure 3
of \cite{Wielgus2022}. This is a result of collimation of radiation by the inflowing fluid at high latitudes. To some extent, the situation resembles radiation beaming in funnels of Polish Doughnuts, discussed in analytic models of acceleration of relativistic jets in quasars and microquasars \citep{1980AcA....30....1J,AbraPiran1980,SikoraWilson1981}. One may also see some analogies with recently calculated models of super-Eddington luminosities observed in ultraluminous X-ray sources (ULXs) \citep{AbarcaParfrey2021,KyanikhooKl2025}. The discovery of coherent pulsations in ULXs showed that some, if not all, ULXs are accreting neutron stars \citep{Bachetti2014}. It seems likely that the ULXs may not be intermediate mass BHs at all, but super critically accreting neutron stars, with their properties compatible with ordinary pulsars and the radiation strongly beamed \citep{KingKluzniakLasota2017, LasotaKing2023}. However, for Polish Doughnuts, the super-Eddington accretion is crucial, since it leads to a very thick torus solution and strong beaming \citep[see also][]{2016A&A...587A..38W}, while for ULXs and puffy discs, the beaming is due to the funnel shape of the photosphere, which is thickened, e.g., by the magnetic field. Therefore, as far as the beaming is concerned, a (vague) analogy with analytic models of the ion tori \citep{Rees1982} seems to be more appropriate.  
The puffy disc may explain some observed properties of real sources better than analytic models, but due to the high computational costs, it does not aspire to replace the models; rather, it highlights several critical discrepancies that should be taken into account when modelling luminous accretion discs. Nonetheless, the simulation results agree qualitatively with the disc-corona model of \cite{Gronki2020}. In future work, we aim to extend the simulations to cover the parameter space in different BH masses, spins, and reach lower mass-accretion rates observed in galactic microquasars as well as in supermassive BHs. 

\section*{Acknowledgements}

The authors thank Andrew Chael and Brandon Curd for their support with the \koral{} code and Ramesh Narayan and Michal Dov\v{c}iak for useful discussions. We gratefully acknowledge the Polish high-performance computing infrastructure PLGrid (HPC Center: ACK Cyfronet AGH) for providing computer facilities and support within the computational grant no. PLG/2025/018611. This work was supported by the Ministry of Education, Youth and Sports of the Czech Republic through the e-INFRA CZ (ID:90254). 

DL acknowledges the Czech Science Foundation (GA\v{C}R) project No. 25-16928O and the internal grant of the Silesian University IGS/17/2024. MA and GT acknowledge the GA\v{C}R grant No. 21-06825X. DL and GT also acknowledge the internal grant of the Silesian University no. SGS/25/2024. MW is supported by a Ramón~y~Cajal grant RYC2023-042988-I from the Spanish Ministry of Science and Innovation and acknowledges financial support from the Severo Ochoa grant CEX2021-001131-S funded by MCIN/AEI/ 10.13039/501100011033. AR is partially supported by the Polish National Science Center grant No. 2021/41/B/ST9/04110. WK acknowledges partial support of this work by the Polish NCN grant No. 2019/35/O/ST9/03965. 

\section*{Data Availability}

 The time- and azimuth-averaged data are published at \url{https://doi.org/10.5281/zenodo.21167643}. The full 3D snapshots from simulations will be provided on request to the corresponding author.



\bibliographystyle{mnras}
\bibliography{bibliography}

\bsp	
\label{lastpage}
\end{document}